\documentclass[reprint,superscriptaddress,showpacs,amsmath,amssymb,prb]{revtex4-2}
\usepackage{natbib}
\usepackage{siunitx}
\usepackage{array}
\usepackage{xcolor}
\usepackage{subfigure} 
\usepackage{amsmath} 
\usepackage{amsfonts}
\usepackage{amssymb}
\usepackage{graphicx}
\usepackage{hyperref}
\usepackage{appendix}
\usepackage{academicons}
\usepackage{orcidlink}
\newcommand{\orcid}[1]{\href{https://orcid.org/#1}{\textcolor[HTML]{A6CE39}{\aiOrcid}}}
\providecommand{\vect}[1]{{\boldsymbol{#1}}}

\usepackage{physics}

\begin{document} 
\title{Strain-induced manipulation of non-collinear antiferromagnets}

\author{Mithuss Tharmalingam} 
\affiliation{Department of Engineering Sciences, University of Agder, 4879 Grimstad, Norway} 
\author{Feodor Svetlanov Konomaev}
\affiliation{Department of Engineering Sciences, University of Agder, 4879 Grimstad, Norway} 
\author{Kjetil M. D. Hals}
\affiliation{Department of Engineering Sciences, University of Agder, 4879 Grimstad, Norway} 
\date{\today}
\newcommand{\Kjetil}[1]{\textcolor{red}{#1}} 
\begin{abstract}
In recent years, there has been growing interest in harnessing non-collinear antiferromagnets (NCAFMs) for applications in antiferromagnetic spintronics. A key requirement for their practical use is the ability to control the spin order in a reliable and tunable manner. In this work, we investigate how the spin order in kagome antiferromagnets --- an important class of NCAFMs --- can be manipulated via strain. Starting from a microscopic spin Hamiltonian, we derive an effective action for the kagome antiferromagnet that captures the coupling between the spin order and the system's strain tensor. At the microscopic level, this coupling arises from strain-induced modifications of the Dzyaloshinskii–Moriya and exchange interactions. Using this effective description, we explore two strain-driven phenomena: (1) strain-induced switching of the antiferromagnetic spin order and (2) the piezomagnetic response. We numerically show that strain facilitates thermally assisted switching between spin configurations of opposite chirality. Specifically, we find that uniform tensile and compressive strain govern both the average switching time and the preferred switching direction between chiral states. Furthermore, we demonstrate that strain induces a net magnetization and provide an experimentally testable prediction of this effect for a typical NCAFM. Our results provide a theoretical framework for modeling strain-induced manipulation of kagome antiferromagnets, underscoring strain as a promising route for functional control of NCAFMs.

\end{abstract}

\maketitle 

\section{Introduction} 
Antiferromagnets have garnered significant interest for spintronic applications due to their potential for ultra-dense device integration and high-speed data processing, surpassing the limitations of conventional ferromagnetic devices~\cite{Jungwirth:np2018,Duine:np2018,Gomonay:np2018,Zelezny:np2018,Nemec:np2018,Libor:np2018, Baltz:rmp2018}. Broadly, antiferromagnets are categorized into two main classes: collinear antiferromagnets (CAFMs) and non-collinear antiferromagnets (NCAFMs)~\cite{Andreev:spu1980}. In CAFMs, the spin arrangement is staggered, with neighboring spins aligned in opposite directions. In contrast, NCAFMs feature a non-collinear configuration of the sublattice spins such that the net spin polarization vanishes in equilibrium. A key consequence of this non-collinear spin order is that NCAFMs are characterized by an SO(3)-valued order parameter field~\cite{Andreev:spu1980}, which leads to richer spin dynamics than in ferromagnets or CAFMs \cite{Han:newton2025}. This complexity makes NCAFMs particularly attractive for antiferromagnetic spintronics, as they combine the advantages of CAFMs with novel emergent phenomena arising from their intricate spin structures. Notably, antiferromagnetic Weyl semimetals~\cite{Kuroda:natmat2017} such as Mn$_3$Sn and Mn$_3$Ge feature electronic band structures with a significant Berry curvature, giving rise to unconventional transport phenomena including the anomalous Hall~\cite{Nakatsuji:nature2015,Nayak:Sadv2016,Kiyohara:pra2015} and Nernst effects~\cite{Ikhlas:nphys2017}.

\begin{figure}[htp]
  \centering
  \begin{subfigure}
    {\includegraphics[scale=0.54]{"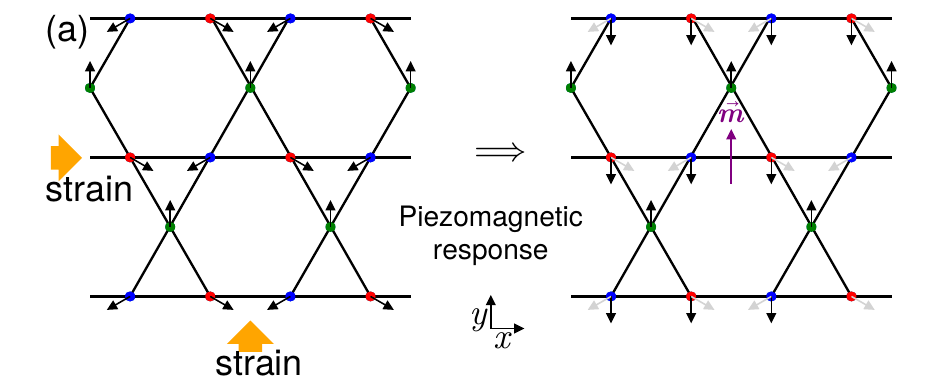"}}
  \end{subfigure}
  \begin{subfigure}
    {\includegraphics[scale=0.54]{"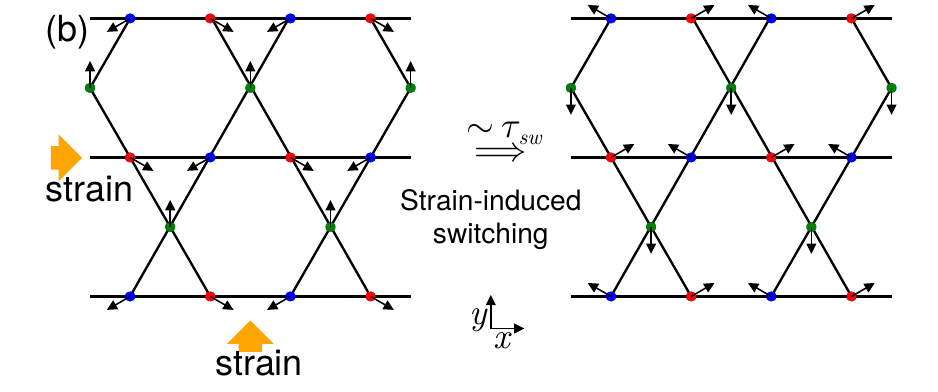"}}
  \end{subfigure}
  \caption{(a) Schematic of piezomagnetism in a kagome antiferromagnet with resulting magnetic moment $\vect{m}$. (b) Schematic of a kagome antiferromagnetic undergoing strain-induced thermally-assisted chirality switching in an average switching time $\tau_{sw}$.}
  \label{fig:strained_kagome_lattice}
\end{figure}

The applicability of NCAFMs relies on effective techniques for manipulating the antiferromagnetic spin order. Various approaches have been explored theoretically, including different types of current-driven torques~\cite{Gomonay:PRB2012, Tserkovnyak:PRB2017, Ochoa:PRB2018, Yamane:PRB2019, Li:PRB2021, Lund:PRL2021, Rodrigues:PRB2022, Lund:PRB2023,Lee:nn2025} as well as frequency-selective spin wave excitations~\cite{Rodrigues:PRL2021}. Experimentally, recent studies on Mn$_3$Sn have demonstrated spin reorientation via current-induced torques~\cite{Takeuchi:nmat2021, Deng:nsr2023, Tsai:nature2020, Higo:Nature2022,Yoon:nmat2023,Xie:nc2022,Yan:am2022} and domain wall writing using laser pulses~\cite{Reichlova:nc2019}. An intriguing yet largely unexplored approach for controlling the spin orientation of NCAFMs involves using strain to modulate the magnetic exchange interactions. Many NCAFMs are expected to have a significant Dzyaloshinskii-Moriya interaction (DMI)—a relativistic exchange interaction that arises from broken spatial inversion symmetry~\cite{Dzyaloshinskii:PR1960, Moriya:PRL2007}. Crucially, the DMI often determines the chirality of the ground-state spin configuration. For instance, in kagome-lattice NCAFMs, the DMI between neighboring spins $\mathbf{S}_i$ and $\mathbf{S}_j$ contains a term that takes the form $D^{z} \hat{\mathbf{z}}\cdot(\mathbf{S}_i \times \mathbf{S}_j )$. When $D^{z}>0$, the combination of DMI and isotropic exchange interaction favors a 120-degree spin configuration with positive chirality, whereas $D^{z}<0$ stabilizes a configuration with negative chirality. First-principles calculations have further revealed that the DMI is highly sensitive to the lattice constant, with its magnitude -- and even its sign -- being tunable via strain~\cite{Meyer:PRB2024, Koretsune:SR2015, Gusev:PRL2020}. This suggests that strain could serve as a promising mechanism for manipulating NCAFMs, for example, by integrating the material with piezoelectric actuators or voltage-controlled substrates that dynamically modify the lattice spacing. Previous investigations of strain effects in NCAFMs have primarily focused on piezomagnetism~\cite{Ikhlas:Nature2022, Meng:nc2024}, magnetostriction~\cite{Meng:nc2024}, and modifications of the excitation spectrum~\cite{Nayga:PRB2022}. However, a theoretical description of the role of DMI in mediating strain-induced changes of the spin order is still missing.

Here, we develop an effective action for strained kagome NCAFMs that captures the coupling between the SO(3) order parameter field and the strain tensor. In our theory, we assume that this coupling originates from the DMI and the exchange interaction, rather than from magnetic anisotropy, which typically gives rise to higher-order terms in the spin-orbit coupling. Furthermore, we retain terms up to second order in both the space-time gradients and the strain tensor.
Based on this formalism, we investigate thermally assisted strain-induced switching and demonstrate that a strain of about 0.25\% is sufficient to initiate a transition between ground states of opposite chirality. Furthermore, we analyze the piezomagnetic response and show that the coupling terms lead to qualitatively distinct behaviors under compressive/tensile versus shear strain. These findings establish strain as a promising and tunable mechanism for voltage-controlled manipulation of NCAFMs.

This paper is organized as follows. In Sec.~\ref{Sec2}, we develop a general effective action description of strained kagome antiferromagnets. Sections \ref{Sec3} and \ref{Sec4} apply this theory to analyze thermally assisted strain-induced switching and the piezomagnetic response, respectively. A discussion is presented in Sec.~\ref{Sec5}, followed by concluding remarks in Sec.~\ref{Sec6}.

\section{General theory} \label{Sec2}
\begin{figure}[h]
  \includegraphics[scale=0.54]{"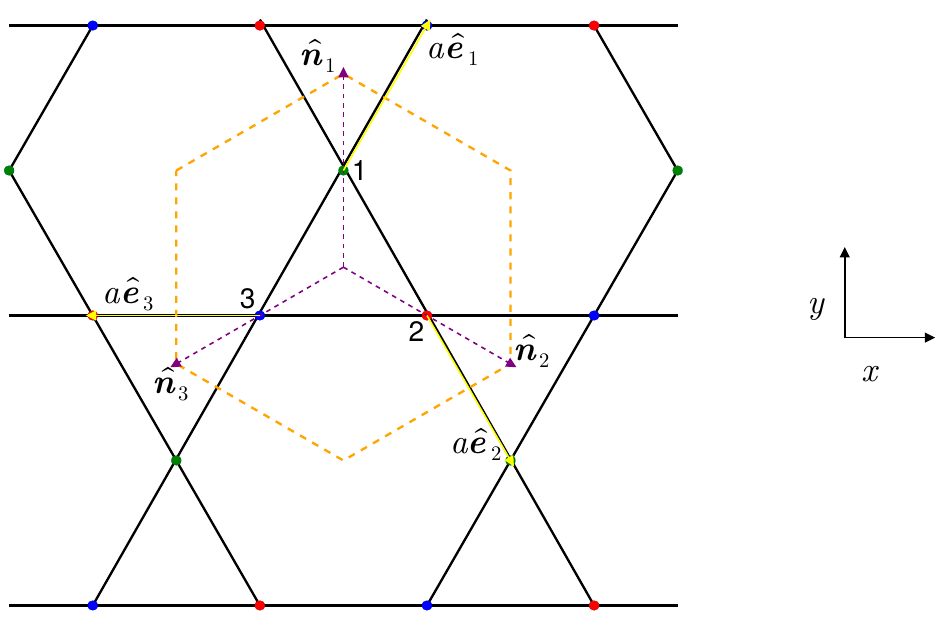"} 
    \caption{The orange dashed lines demarcate the magnetic unit cell. The easy axes vectors $\hat{\mathbf{n}}_i$ lie in the plane, and the vector $a\hat{\mathbf{e}}_{i}$ connects lattice point $i$ to its nearest neighbour that contributes to the energy of one unit cell (see Sec.~\ref{sec:effectiveactionkagome}). The $z$-axis is perpendicular to the plane of the kagome lattice.}
  \label{fig:unit_cell}
\end{figure}

We consider a strained two-dimensional kagome antiferromagnet (see Fig. \ref{fig:strained_kagome_lattice}). Our main aim is, starting from a spin Hamiltonian, to derive an action that captures the coupling to the strain tensor induced by the DMI and exchange interaction. The kagome antiferromagnet is modeled by the following spin Hamiltonian \cite{Rodrigues:PRB2022,Ulloa:PRB2016},
\begin{align}\label{eq:hamiltonian}
    \mathcal{H}
        =&\sum_{\langle ij\rangle}J_{ij} \mathbf{S}_i\cdot \mathbf{S}_j + \sum_i \left[K_\perp (\mathbf{S}_i\cdot \hat{\mathbf{z}})^2-K(\mathbf{S}_i\cdot \hat{\mathbf{n}}_i)^2\right]\notag \\
        &+ \sum_{\langle ij\rangle}\mathbf{D}_{ij}\cdot(\mathbf{S}_i \times \mathbf{S}_j ) - \sum_i \mathbf{h}_i \cdot \mathbf{S}_i.
\end{align}
The Hamiltonian \eqref{eq:hamiltonian} contains five different contributions. The first term describes the isotropic exchange interaction, characterized by the exchange constant $J_{ij}$, between nearest neighbour spins at sites $\mathbf{r}_i = \mathbf{r}_i^{(0)}+ \mathbf{u}(\mathbf{r}_i)$ and $\mathbf{r}_j= \mathbf{r}_j^{(0)}+ \mathbf{u}(\mathbf{r}_j)$, where $\mathbf{r}_i^{(0)}$ is the equilibrium position of the spin and $\mathbf{u}(\mathbf{r}_i)$ is the displacement field. The second term describes an easy-plane anisotropy (perpendicular to $z$) characterized by the anisotropy constant $K_\perp>0$. The third term, characterized by an anisotropy constant $K>0$, represents an in-plane anisotropy that favors spin orientations either parallel or antiparallel with the in-plane anisotropy vectors $\hat{\mathbf{n}}_i$.  
Here, the in-plane anisotropy vectors at the three spin sublattices are $\hat{\mathbf{n}}_1 = [0,1,0]$, $\hat{\mathbf{n}}_2 = [\sqrt{3}/2,-1/2,0]$, and $\hat{\mathbf{n}}_3 =  [-\sqrt{3}/2,-1/2,0 ]$.
The fourth term is the DMI~\cite{Dzyaloshinskii:PR1960,Moriya:PRL2007}, parametrized by the DMI vector $\mathbf{D}_{ij}$ between the nearest neighbour spins at sites $\mathbf{r}_i$ and $\mathbf{r}_j$. The DMI vector has both an out-of-plane $D^z_{ij}= D^z$ and an in-plane $D^\parallel_{ij}= D^\parallel$ component \cite{Willis:PRB2001,Rodrigues:PRB2022}:
\begin{align}
  \mathbf{D}_{13} 
    & = D^z_{13} \hat{\mathbf{z}} + (\hat{\mathbf{e}}_{1}\times \hat{\mathbf{z}}) D^\parallel_{13} , \\
  \mathbf{D}_{21} 
    & = D^z_{21} \hat{\mathbf{z}} + (\hat{\mathbf{e}}_{2}\times \hat{\mathbf{z}}) D^\parallel_{21} , \\
  \mathbf{D}_{32} 
    & = D^z_{32} \hat{\mathbf{z}} + (\hat{\mathbf{e}}_{3}\times \hat{\mathbf{z}}) D^\parallel_{32} .
\end{align}
Here $\hat{\mathbf{e}}_{1}= [1/2, \sqrt{3}/2, 0]$, $\hat{\mathbf{e}}_{2}=[1/2, -\sqrt{3}/2,0]$ and $\hat{\mathbf{e}}_{3}=[-1,0,0]$ are unit vectors connecting lattice site $i$ to its nearest neighbor that contributes to the energy of a single unit cell (see Fig. \ref{fig:unit_cell} and Sec.~\ref{sec:effectiveactionkagome}).
Finally, we consider multiplicative noise in the form of a thermal white noise field $\mathbf{h }$  satisfying the fluctuation-dissipation theorem,
\begin{align}
  \langle h_{\alpha} (t)h_{\beta}(t')\rangle = \frac{2\hbar k_B T \alpha_G }{S} \delta_{\alpha\beta}\delta(t-t'),
\end{align}
and $\langle \mathbf{h } \rangle =0$, where $\alpha_G$ is the Gilbert damping parameter, $S$ the spin, and $\langle \dots \rangle $ the statistical average.
The white noise field microscopically arises from modulations in the local magnetic anisotropy and exchange interactions induced by lattice vibrations. Since these modulations typically occur on the scale of one unit cell, we assume that the noise field is uniform within each unit cell and uncorrelated between different cells.

The ground state of the Hamiltonian~\eqref{eq:hamiltonian} depends on whether the ratio $D^z/K$ is less or greater than $1/(4\sqrt{3})$ \cite{Lund:PRB2023}. When $D^z/K<1/(4\sqrt{3})$, 
the spins are oriented in the same or opposite direction to the in-plane vectors $\hat{\mathbf{n}}_i$. We refer to this state as the $(+)$-chiral state. Conversely, when $D^z/K>1/(4\sqrt{3})$, the ground state attains a spin configuration with negative chirality (referred to as the $(-)$-chiral state), which is related to the $(+)$-chiral state by a reflection about the $xz$-plane (see Fig. \ref{fig:strained_kagome_lattice} (b)). 

It is known that the DMI and the exchange interaction are heavily dependent on strain \cite{Gusev:PRL2020,Nayga:PRB2022,Meyer:PRB2024,Koretsune:SR2015}. We incorporate strain into our theory by allowing the DMI vector and the exchange parameter to depend on the lattice spacing and subsequently expanding them about the equilibrium lattice constant $a$:
\begin{align}
  J_{ij}
    & =J(a)+A_{ij}^{\alpha\beta}\varepsilon_{\alpha\beta}; \label{eq:exchange_parameter}\\
  D^z_{ij}
    & = D^z(a) + B_{ij}^{\alpha\beta} \varepsilon_{\alpha\beta};\label{eq:DMIz}\\ 
  D^\parallel_{ij}
    & = D^\parallel(a) + C_{ij}^{\alpha\beta} \varepsilon_{\alpha\beta}\label{eq:DMIparallel}.
\end{align} 
Here, the coupling tensors are defined as $A_{ij}^{\alpha\beta}=a \left(\frac{\partial J }{\partial a}\right) e_{i,\alpha} e_{i,\beta}$, $B_{ij}^{\alpha\beta} =a  \left( \frac{\partial D^z_{ij  }}{\partial a} \right) e_{i,\alpha} e_{i,\beta}$, $C_{ij}^{\alpha\beta} =a  \left( \frac{\partial D^\parallel_{ij}}{\partial a} \right) e_{i,\alpha} e_{i,\beta}$, and $\varepsilon_{\alpha\beta} = \frac{1}{2} \left( \frac{\partial {u}_\alpha(\mathbf{r}_i)}{\partial r_\beta} + \frac{\partial {u}_\beta(\mathbf{r}_i)}{\partial r_\alpha} \right)$ is the strain tensor (where $\alpha = x,y,z$). Throughout, Einstein’s summation convention is implied for repeated indices.

The spin dynamics are governed by the spin action $\mathcal{ S}=\sum_i \hbar \int \text{d}t \mathbf{A}(\mathbf{S}_i)\cdot \dot{\mathbf{S}}_i - \int \text{d} t \mathcal{H}$ and dissipation functional $\mathcal{G}=\sum_i\hbar \int \text{d}t\left[(\alpha_G /2)\dot{\mathbf{S}}^2_i\right]$ \cite{Dombre:PRB1989,Lund:PRL2021}, where the vector potential $\mathbf{A}$ in the kinetic part of the action is defined via $\nabla \times A(\mathbf{S}_i)=\mathbf{S}_i/S$. 
To derive an effective description of the NCAFM, we express each spin $\mathbf{S}_i$ in terms of the rotation matrix $\mathbf{R} \in \text{SO}(3)$ and the canting field vector $\mathbf{L}$ \cite{Dombre:PRB1989},
\begin{align}\label{eq:spinncaf}
  \mathbf{S}_i
      & = \frac{S\mathbf{R} (\hat{\mathbf{n}}_i + a \mathbf{L})}{\sqrt{1+2a \hat{\mathbf{n}}_i \cdot \mathbf{L} + a^2 \mathbf{L}^2}}, \hspace{1cm} i \in \{ 1,2,3 \} .
\end{align}
In what follows, we represent the rotation matrix in terms of nautical angles,  $\mathbf{R}(\theta,\phi,\psi)=\mathbf{R}_z(\theta)\mathbf{R}_y(\phi)\mathbf{R}_x(\psi)$,  where $\psi (t), \phi(t)$ and $\theta(t)$ denote the angle of rotation about the $x$-, $y$- and $z$-axis, respectively.   
Together, $\mathbf{R}$ and $\mathbf{L}$ provide the necessary six degrees of freedom for parametrizing the orientation of the three sublattice spins within one unit cell -- three from $\mathbf{R}$ through the nautical angles and three from $\mathbf{L}$. Furthermore, we assume that the canting vector $a \mathbf{L}$ is a small quantity. 

Substituting Eq.~\eqref{eq:spinncaf} along with Eqs.~\eqref{eq:exchange_parameter}-\eqref{eq:DMIparallel} into the spin Hamiltonian~\eqref{eq:hamiltonian}, followed an expansion to second order in the space-time gradients, we obtain the action $\mathcal{S}= \int \text{d}t \text{d}\mathbf{r} \mathcal{L}$ in the continuum limit, where the Lagrangian density is
$\mathcal{L}= \mathcal{T} - \mathcal{F}_e - \mathcal{F}_a- \mathcal{F}_{\text{DMI}}-\mathcal{F}_h- \mathcal{F}_s.$  
To second order in the frequency, the kinetic part $\mathcal{T}$ is
\begin{align}
  \mathcal{T}
    & = K_T\mathbf{m}\cdot \mathbf{V},
\end{align}
where the vector ${V}_\alpha=-\frac{1}{2}\epsilon_{\alpha\beta\gamma}(R^{-1}\partial_tR)_{\beta\gamma}$ is determined by the time derivative of the rotation matrix. 
The vector field $\mathbf{m}=\mathbf{T} \mathbf{L}$ relates to the magnetization via $\mathbf{M}=\mathcal{N} K_M\mathbf{R} \mathbf{m}$, where $\mathcal{N}$ is the number of kagome planes contained in the 3D unit cell, and $\mathbf{T}$ is a diagonal matrix described by $2T_{xx} = 2T_{yy} = T_{zz} = 1$. The constants $K_T$ and $K_M$ are defined in Table~\ref{tab:1}. To second order in spatial gradients, $\mathbf{m}$, and $\epsilon_{\alpha\beta}$, the energy densities become
\begin{align}
  \mathcal{F}_e 
    & = K_e\mathbf{m}^2 - \Lambda_{mm'}^{\alpha\beta}\left[(\partial_\alpha R^T)(\partial_\beta R)\right]_{mm'}, \label{Eq:Fe} \\
  \mathcal{F}_\text{a}
    & = -R_{\alpha\beta}R_{\alpha'\beta'} (K^{(1)}_{\alpha\beta\alpha'\beta'}+m_\gamma K^{(2)}_{\alpha\beta\gamma\alpha'\beta'}) \notag \\
    & \quad -R_{\alpha\beta}R_{\alpha'\beta'}m_\gamma m_{\gamma'} K^{(3)}_{\alpha\beta\gamma\alpha'\beta'\gamma'},\\
  \mathcal{F}_{\text{DMI}}
    & = J_{\beta\beta';\gamma\gamma'}^{ab}(\partial_a R_{\beta\beta'})(\partial_b R_{\gamma\gamma'})\notag \\
    & \quad+ K_{\beta\beta';\gamma\gamma'}^DR_{\beta\beta'}R_{\gamma\gamma'}- \mathbf{H}_D\cdot \mathbf{m},\\
  \mathcal{F}_h
    & = -K_h  h_\beta  R_{\beta \alpha} m_\alpha , \label{Eq:Fh}\\
  \mathcal{F}_s
    & = C_{\alpha\beta \gamma}\varepsilon_{\alpha\beta} m_\gamma + {D}^{(1)\alpha'\beta'}_{\alpha\beta}\varepsilon_{\alpha'\beta'}  R_{\alpha\beta} \notag\\
    & \quad  +{D}^{(2)\alpha'\beta'}_{\alpha\beta\gamma}\varepsilon_{\alpha'\beta'} R_{\alpha\beta} m_\gamma . \label{Eq:Fs}
\end{align}
Here, $\mathcal{F}_e$, $\mathcal{F}_a$, and $\mathcal{F}_{\rm DMI}$ represent the exchange, anisotropy, and DMI energy densities, respectively.  $\mathcal{F}_h$ describes the energy density due to the stochastic white noise field, while  $\mathcal{F}_s$ determines the energy density produced by the strain. The first term in $\mathcal{F}_s$ originates from the exchange energy, whereas the last two arises from the strain-induced modifications of the DMI. 
The details of the above derivation, along with definitions of all constants and tensors, are provided in Appendix \ref{sec:effectiveactionkagome}.
Note that strain also alters the lattice energy. However, since we consider only static strain and disregard lattice dynamics due to phonons, the lattice energy contribution is omitted from our analysis.

The dissipation functional in terms of the nautical angles reads~\cite{Lund:PRL2021}
\begin{align}
  \mathcal{G}
    & = \frac{\alpha }{2}\int {\rm d} t {\rm d} \mathbf{r} \left(\dot{\psi}^2 + \dot{\phi}^2 + \dot{\theta}^2 - 2 \dot{\theta}\dot{\psi}\sin \phi \right) , 
\end{align}
where $\alpha= 3\hbar\alpha_GS^2/V_{uc}$ is determined by the Gilbert damping $\alpha_G$ of the spins ($V_{uc}$ is the unit cell volume). 
The equations of motion for the nautical angles and $\mathbf{m}$ are found from the variational equation 
\begin{align}\label{eq:fullaction}
  \frac{\delta \mathcal{S} }{\delta \rho} 
    & = \frac{\delta \mathcal{G}}{\delta \dot{\rho}}, \hspace{0.8cm} \rho\in \left\{ m_x,m_y,m_z,\theta,\phi,\psi \right\}, 
\end{align}
which we will use in the following for investigating thermally assisted strain-induced switching.

\section{Numerical investigation of strain-induced switching}\label{Sec3}
\begin{figure}[h]
  \includegraphics[scale=0.54]{"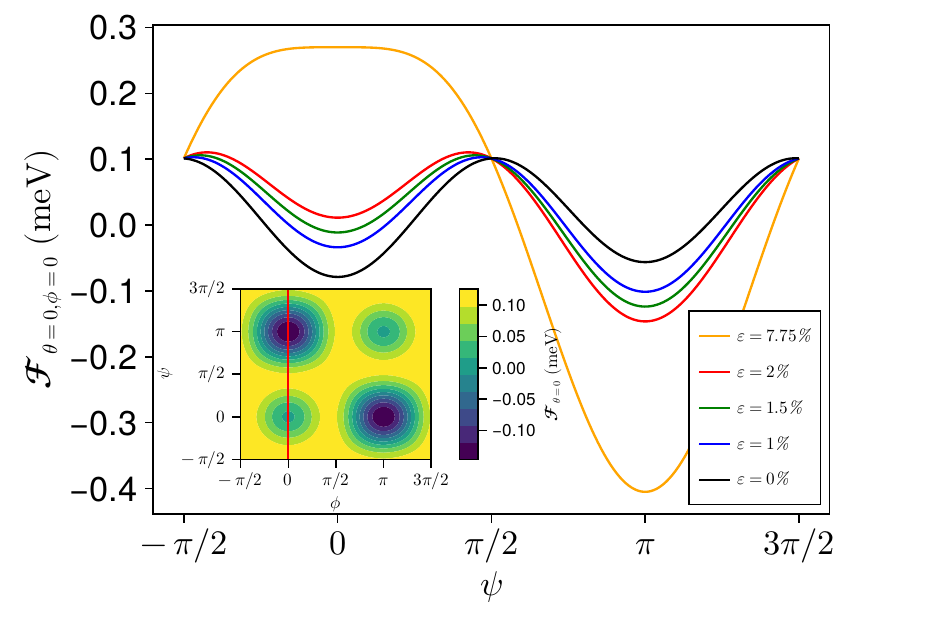"} 
  \caption{Analytic plot of the free energy at 7.75\%, 2\%, 1.5\%, 1\% and 0\% uniform strain while putting the $\theta$, $\phi$ and magnetization to zero. Applying more strain causes the potential barrier to shrink, until it completely vanishes at 7.75\% uniform tensile strain. The inset contains the free energy at $\varepsilon = 2\%$ while only putting $\theta$ and the magnetization to zero. The red line shows the slice of the main plot of this figure.  The chosen model coefficients are: $J = 10$ meV, $K=0.03$ meV, $K_{\bot}=0.09$ meV, $D^z= K/ 8\sqrt{3}$, $(a/D^z)\frac{\partial D^z }{\partial a }= 400 $, $S=1.0$, $\alpha_G=0.01$ \cite{Lund:PRB2023,Gusev:PRL2020}}
  \label{fig:free_energy}
\end{figure} 

\begin{figure*}[htp]
  \centering
  \begin{subfigure}
      {\includegraphics[scale=0.54]{"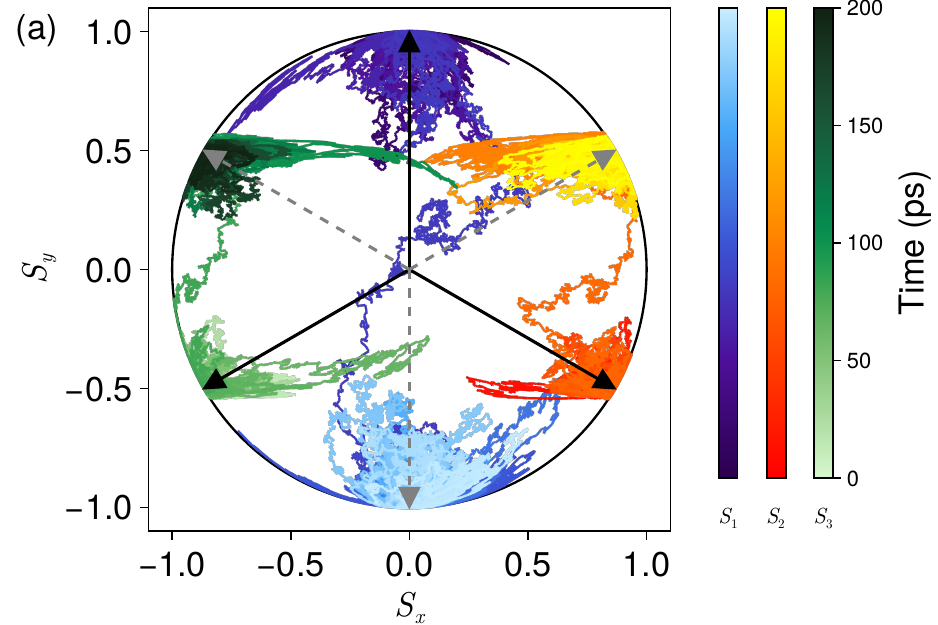"}}
  \end{subfigure}\quad
  \begin{subfigure}
      {\includegraphics[scale=0.54]{"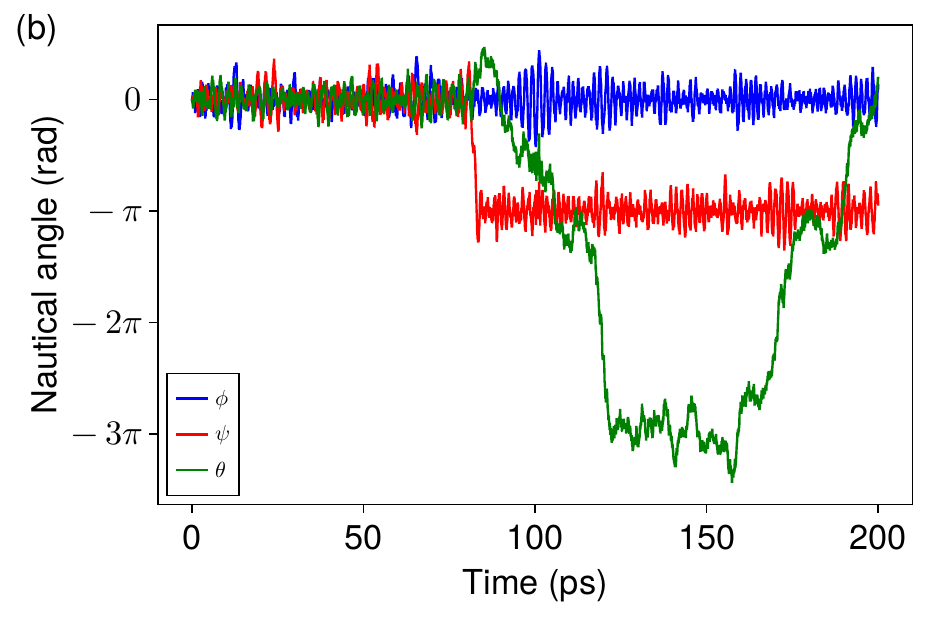"}}
  \end{subfigure}
  \caption{(a) Two-dimensional projection of a representative macrospin trajectory at 2\% strain and $T=35$~$\mathrm{K}$, obtained by numerically integrating the six Stratonovich SDEs governing the dynamics of the nautical angles and $\mathbf{m}$. The model parameters are identical to those used in Fig.~\ref{fig:free_energy}. Black arrows indicate the $(+)$-chiral state, while dashed gray arrows represent the $(-)$-chiral state. (b) Evolution of the nautical angles corresponding to the same trajectory.}
  \label{fig:macrospin_trajectory}
\end{figure*}
As previously mentioned, the ground state depends on whether the ratio $D^z/K$ is less than or greater than $1/(4\sqrt{3})$. In the following, we aim to tune $D^z$  such that the system transitions from the $(+)$-chiral state to the $(-)$-chiral state.  To achieve this, we employ strain to modify $D^z$ as described in Eq.~\eqref{eq:DMIz}. As illustrated in Fig.~\ref{fig:free_energy}, this results in a noticeable change of the energy landscape. At 2\% uniform strain, the $(+)$-chiral state remains metastable. Consequently, the system must overcome an energy barrier to reach the $(-)$-chiral state. At finite temperature, the antiferromagnet can escape from the local minimum associated with the $(+)$-chiral state through thermal excitations. In our theoretical model, the effects of thermal excitations are captured by the white noise field in Eq.~\eqref{Eq:Fh}.   

To investigate the dynamical evolution from the $(+)$- to the $(-)$-chiral state we derive the six coupled equations of motions for the nautical angles and $\mathbf{m}$ from Eq.~\eqref{eq:fullaction},  
starting with a purely antiferromagnetic system with $[m_x,m_y,m_z]=[0,0,0]$. Since $D^\parallel$ induces a weak ferromagnetic phase, we set $D^\parallel = 0$. 
Additionally, we consider a uniform antiferromagnet, neglecting spatial variations in the order parameter field $\mathbf{R}$.
This corresponds to a macrospin approximation, where the antiferromagnetic configuration is represented by three macrospins parametrizing the orientation of the three sublattice spins.
The expressions for energy densities used in Eq.~\eqref{eq:fullaction} are given in Appendix~\ref{sec:fullaction}.  
We numerically solve the resulting stochastic differential equations (SDE), assuming that the evolution represents a Markov process. Moreover, the process evolves continuously over time making it a Wiener process. As a result, we obtain six Stratonovich SDEs which we solve in Mathematica using the $\mathtt{StratonovichProcess}$ function. In Mathematica, we get an approximate numerical solution using the Euler-Maruyama method. 

An example of a macrospin trajectory for a solution displaying spin switching is given in Fig.~\ref{fig:macrospin_trajectory}a in a system with a uniform strain of $\varepsilon = \varepsilon_{xx} = \varepsilon_{yy}$ and $\varepsilon_{xy} = \varepsilon_{yx} = 0$, of $\varepsilon = 2\%$. The model coefficients are given in the caption and have been chosen similarly to Ref. \cite{Lund:PRB2023}. $\frac{\partial D^z}{\partial a}$ has been chosen in accordance to Ref. \cite{Gusev:PRL2020}. For simplicity, we put $\frac{\partial J}{\partial a}$ to zero. 
This approximation is justified when spatial variations are neglected and the exchange interaction only couples to the spin system through a small modification of the $\mathbf{m}$ field. Moreover, the potential barrier between the $+$ and $-$ chiral states (as plotted in Fig.~\ref{fig:free_energy}) is unaffected by $\frac{\partial J}{\partial a}$. 

The trajectory starts by wandering around its starting point (solid black arrows) in a $(+)$-chiral state and, after some time, it reaches the potential-barrier from which it quickly descends to a $(-)$-chiral state (dashed gray arrows). 
The switching of chirality corresponds to the nautical angle $\psi$  changing from 0 to $\pi$. It is also possible for chirality to switch via the nautical angle $\phi$ undergoing the same change, though this occurs less frequently. 
This asymmetry can be attributed to the flat band structure of spin waves associated with rotations about the 
$x$-axis \cite{Lund:PRL2021}, which implies weak spin stiffness and thus facilitates switching through the $\psi$ angle. 
In Fig.~\ref{fig:macrospin_trajectory}a we have suppressed the $\theta$ dependence to see the spin switching more clearly. After switching to the $(-)$-chiral state, the $\theta$ angle does not settle to a constant value but instead varies with time (see Fig.~\ref{fig:macrospin_trajectory}b, where the same trajectory is presented using nautical angles).  This is a consequence of that the ($-$)-chiral state exhibits gapless excitations~\cite{Lund:PRB2023} corresponding to uniform rotations about the $z$-axis. 
Switching back to the $(+)$-chiral state is easily achieved by removing the strain.

\begin{figure}[h]
  \includegraphics[scale=0.54]{"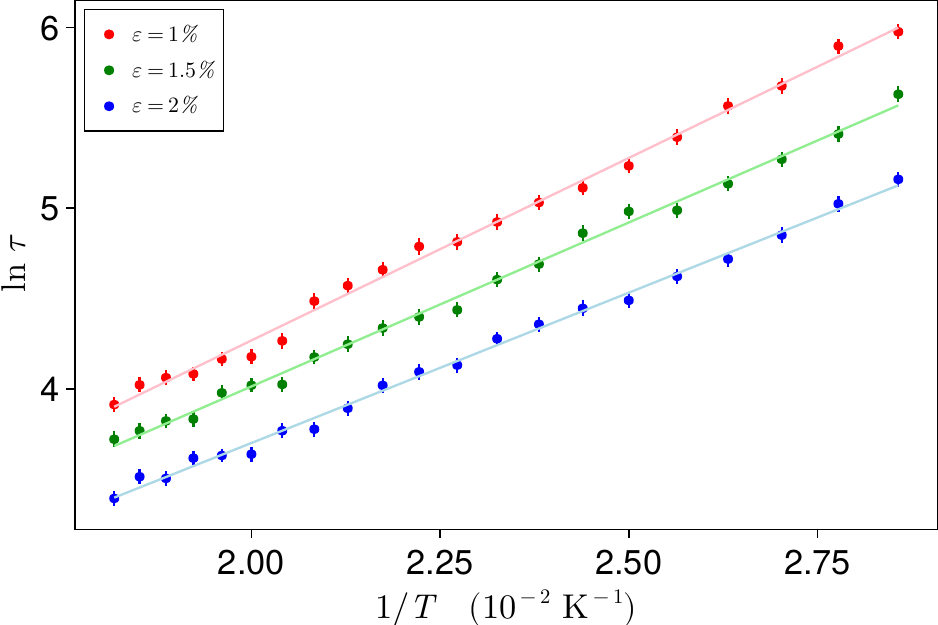"}
  \caption{Logarithm of the average switching times from the $(+)$-chiral state to the $(-)$-chiral state as a function of the inverse temperature for different strains. The lines through the data are linear fits to the data. The numerical data is collected from a set of 500 macrospin systems with the same model coefficients as in Fig. \ref{fig:free_energy}.}
  \label{fig:switching_times}
\end{figure}

We have repeated these simulations for a set of 500 realizations of the macrospin system to find the average spin switching times $\tau$ as a function of the temperature $T$ at different strengths of uniform strain $\varepsilon$. The results are given in Fig. \ref{fig:switching_times} and suggest that the switching times follow the functional form $\tau = A \exp(C \frac{\Delta E}{k_B T})$, where $A$ is the switching time at high temperatures. The dimensionless quantity $C$, along with the potential barrier $\Delta E$ between the (+) and (-) chiral states, determines the slope of $\rm ln (\tau)$. As we can see, the switching time decreases as we increase the uniform tensile strain. This behaviour is expected when looking at the free energy, as the energy barrier decreases as we increase strain. At a strain of 0.25\%, the free energy of the $(+)$-chiral state and the $(-)$-chiral state are equal. This means that for larger strains, for example $\varepsilon = 2\%$, the $(-)$-chiral state is favored, as seen from Fig. \ref{fig:free_energy}.  At a strain of 7.75\%, the barrier vanishes completely, and switching will even occur in the limit of vanishing thermal fluctuations.
\section{Piezomagnetic response}\label{Sec4}
To explore the piezomagnetic response in the $(+)$-chiral and $(-)$-chiral state -- characterized by the nautical angles $(\theta,\phi,\psi)=(0,0,0)$ and $(\theta,\phi,\psi)=(0,0,\pi)$, respectively -- 
we minimize the effective action with respect to  $\mathbf{m}$ in each state.  This is done by solving the variational equation $\frac{\delta \mathcal{S}}{\delta \mathbf{m}}={0}$, yielding the following expression for the strain-induced magnetization in a static, homogeneous system 
to leading order in $D^z/J$ and $K/J$:
\begin{align}
  \mathbf{M}^{(+)}
    & = \frac{\mathcal{N}K_M }{2V_{uc}K_e}\Big[a (E_s^{(1)}+E_s^{(2)})\varepsilon_{xy} \hat{\mathbf{x}}\notag\\
    &\quad +\frac{a }{2} (E_s^{(1)}+E_s^{(2)})(\varepsilon_{xx}-\varepsilon_{yy}) \hat{\mathbf{y}}\Big];\label{eq:magnetization_plus}\\
  \mathbf{M}^{(-)}
    & = \frac{\mathcal{N} K_M }{2V_{uc}K_e}\Big[ a (E_s^{(1)}-E_s^{(2)})\varepsilon_{xy} \hat{\mathbf{x}}\notag\\
    &\quad +\frac{a }{2} (E_s^{(1)}-E_s^{(2)})(\varepsilon_{xx}-\varepsilon_{yy}) \hat{\mathbf{y}} - K_{a1}^{(-)} \hat{\mathbf{y}}\Big].\label{eq:magnetization_minus}
\end{align}
Two notable phenomena arise in this context. First, in the absence of strain, the magnetization in the $(-)$-chiral state, $\mathbf{M}^{(-)}$, exhibits a nonzero y-component, in contrast to the zero magnetization of the (+)-chiral state, $\mathbf{M}^{(+)}$. The non-vanishing magnetization of the $(-)$ chiral state originates from the magnetic anisotropy, which causes the spins to align preferentially along the easy axes, resulting in a tilt toward the $-y$ direction. This feature could potentially be used to detect chirality switching. Second, under non-uniform strain -- i.e., when $\varepsilon_{xx}\neq \epsilon_{yy}$ and possibly $\varepsilon_{xy}=\varepsilon_{yx}\neq 0$ -- the magnetization in both chiral states becomes strain-dependent. This well-known phenomenon, known as the piezomagnetic effect, has previously been experimentally observed in kagome antiferromagnets \cite{Ikhlas:Nature2022}.

\section{Discussion}\label{Sec5}
Thermal fluctuations can trigger chirality switching, as demonstrated in our numerical simulations. Experimentally, this switching can be detected by measuring the finite magnetization of the $(-)$-chiral state in an unstrained system. To estimate this equilibrium magnetization, we use typical material parameters for  $\text{Mn}_3\text{Sn}$: $2a=5.665$~Å, out-of-plane lattice constant $c=4.536$~Å, spin $S=1$, gyromagnetic ratio $\gamma= 1.76\times 10^{11}$~(Ts)$^{-1}$, $J=17.53~\mathrm{meV}$, $D^z=0.833~\mathrm{meV}$ and $K=0.196~\mathrm{meV}$ \cite{Higo:Nature2022, Tomita:JPSJ2020}.  
$V_{uc}= 2\sqrt{3} a^2 c $ is the volume of the 3D unit cell containing $\mathcal{N}=2$ kagome planes.
With these values, we find a measurable magnetization of approximately $\abs{\mathbf{M}^{(-)}} = 1.6 \times 10^{3}~\text{A} \text{m}^{-1}$. 

Using the material parameters given above, we can also estimate the piezomagnetic response under non-uniform strain. For example, assuming 
$(\varepsilon_{yy} - \varepsilon_{xx})/2 =0.1\% $ or a shear strain of $\varepsilon_{xy}= 0.1\%$ yields the strain-induced magnetizations $\abs{\mathbf{M}^{(+)}}=(2.8-8.9)\times 10^{3}\  \text{A} \text{m}^{-1}$ and $\abs{\mathbf{M}^{(-)}}=1.6\times 10^{2}-6.2\times 10^{3}\  \text{A} \text{m}^{-1}$ in the two ground states with $\pm$ chirality. Here, the range $\frac{\partial D^z }{\partial a } = (1.8-9.2) \times 10^{11}\ \mathrm{meV} \text{m}^{-1}$ and $\frac{\partial J }{\partial a } = 9.3\times 10^{10} \mathrm{meV} \text{m}^{-1}$ have been chosen in accordance to first principles calculations in Ref. \cite{Gusev:PRL2020,Khmelevskyi:PRB2016}. 
While not specific to $\text{Mn}_3\text{Sn}$, these values provide a reasonable order-of-magnitude estimate. 
To get a better estimation, explicit experimental data is needed on the strain dependence of the DMI and the exchange parameter in kagome antiferromagnets, such as iron jarosites and $\text{Mn}_3X$ with $X = \text{Sn, Ga, Ge}$. 


\section{Conclusion}\label{Sec6}
To conclude, we have calculated the action of a strained kagome antiferromagnet. Our formalism shows that strain can drive a transition in the ground state from the  $(+)$-chiral configuration to the $(-)$-chiral one and vice versa. Thermal fluctuations can then trigger magnetic state switching, which we demonstrated numerically. In particular, increasing uniform tensile strain reduces the spin switching time by lowering the energy barrier between the two chiral states. This switching could potentially be detected experimentally, as the $(-)$-chiral state exhibits a small but finite equilibrium magnetization, in contrast to the zero magnetization of the $(+)$-chiral state. Finally, we examined how various strain profiles generate distinct piezomagnetic responses and provided quantitative estimates of these effects using typical material parameters for NCAFMs.

\section{Acknowledgements}
We thank Mathias Kläui for stimulating discussions.
KMDH acknowledges funding from the Research Council of Norway via Project No. 334202.


\appendix
\section{Derivation of action}\label{sec:effectiveactionkagome}
In this appendix, we derive the action in the continuum limit of a kagome antiferromagnet subjected to strain. 

We recall that due to the fact that the exchange energy is large, $a\mathbf{L}$ is a small quantity, which allows us to write Eq.~\eqref{eq:spinncaf} as 
\begin{align}\label{eq:spin_approximation}
  \mathbf{S}_i 
    & \simeq S \mathbf{R} \left(\hat{\mathbf{n}}_i + \mathbf{\Delta}_i\right), \hspace{1 cm} (\mathbf{\Delta}_i \cdot \hat{\mathbf{n}} = 0),
\end{align}
where $\mathbf{\Delta}_i = a (\mathbf{L}-(\hat{\mathbf{n}}_i\cdot \mathbf{L})\hat{\mathbf{n}}_i)$. Using this we find,
\begin{align}\label{eq:total_spin_polarization}
  \sum_i \mathbf{S}_i 
    & = 3 a S \mathbf{R} \mathbf{m},
\end{align}
for the total spin polarization of each unit cell. 
Here, $\mathbf{m}= \mathbf{T} \mathbf{L}$ where $\mathbf{T}$ is a diagonal matrix with $2T_{xx}= 2T_{yy}=  T_{zz}= 1$.
The rotation matrix $\mathbf{R}$ serves as the order parameter of the NCAFMs, while $\mathbf{m}$ characterizes the local magnetization. Both $\mathbf{R}$ and $\mathbf{m}$ are constant within a single unit cell and are assumed to vary smoothly in space and time. In the following, we express each term of the spin Hamiltonian in terms of $\mathbf{R}$ and $\mathbf{m}$. 
To derive a continuum model for the action we first evaluate the energy $H^{(l)}$ of a single unit cell $l$, then sum over all unit cells, and take the continuum limit as $\sum_l H^{(l)} \rightarrow \int {\rm d}\mathbf{r} H^{(l)}/V_{uc}$ where $\mathcal{F}= H^{(l)}/V_{uc}$ yields the energy densities
in Eqs.~\eqref{Eq:Fe}-\eqref{Eq:Fs}. 

\begin{table}
  \centering
    \caption{\label{tab:1} Constants definitions}
  \begin{tabular}[t]{lc}
  \hline  \\ [-2.0ex]
  Symbol &   Expression \\
  \hline  \\ [-1.5ex] \vspace{1mm}  
  $K_M $ & $\frac{3a \gamma \hbar S}{V_{uc}}$ \\  \vspace{1mm}
  $K_T $ & $\frac{3a\hbar S}{V_{uc}}$ \\  \vspace{1mm}
  $K_e$ & $\frac{9a^2S^2J}{V_{uc}} $  \\ \vspace{1mm}
  $K_h$ & $\frac{3 a S}{V_{uc}} $  \\ \vspace{1mm}
  $K_{a1}^{(-)}$ &  $3aS^2K$ \\ \vspace{1mm} 
  $K_d$ &  $3\sqrt{3}S^2D^z$ \\ \vspace{1mm} 
  $E_s^{(1)}$ &  $9aS^2 \frac{\partial J}{\partial a}$ \\ \vspace{1mm} 
  $E_s^{(2)}$ &  $3\sqrt{3}aS^2\frac{\partial D^z}{\partial a} $ \\ 
  \hline
  \end{tabular}
\end{table} 
\begin{table}
  \centering
    \caption{\label{tab:2} Tensors definitions where the nearest-neighbour sums are over $\langle i,j\rangle \in \{ \langle 1,3\rangle, \langle 2,1\rangle , \langle 3,2\rangle \}$}
  \begin{tabular}[t]{lc}
  \hline  \\ [-2.0ex]
  Symbol &   Expression \\
  \hline  \\ [-1.5ex] \vspace{1mm}  
  $\Lambda_{\gamma\gamma'}^{\alpha\beta}$ & $\frac{a^2S^2J}{V_{uc}}\sum\limits_{\langle i,j\rangle }e_{i\alpha}e_{i\beta}n_{i\gamma}n_{j\gamma'}$  \\ \vspace{1mm}
  $k^{iz}_{\alpha,\alpha'}$ & $Kn_{i\alpha}n_{i\alpha'}-K_{\bot} \delta_{z\alpha}\delta_{z\alpha'}$  \\ \vspace{1mm}
  $K^{(1)}_{\alpha\beta\alpha'\beta'}$ &   $\sum\limits_{i=1,2,3} \frac{S^2}{V_{uc}}k^{iz}_{\alpha,\alpha'}n_{i\beta}n_{i\beta'} 
  $\\  \vspace{1mm}
  $K^{(2)}_{\alpha\beta\gamma\alpha'\beta'}$ &  $\sum\limits_{i=1,2,3} \frac{2S^2 a}{V_{uc}} k^{iz}_{\alpha,\alpha'}[n_{i\beta}\delta_{\beta'\gamma'}-n_{i\beta}n_{i\beta'} n_{i\gamma'}]T^{-1}_{\gamma'\gamma}$  \\  \vspace{1mm}
  $K^{(3)}_{\alpha\beta\gamma\alpha'\beta'\gamma'}$  & \begin{tabular}[t]{@{}c@{}}$ \sum\limits_{i=1,2,3} \frac{a^2 S^2}{V_{uc}}k^{iz}_{\alpha,\alpha'}[   \delta_{\beta\gamma''} +  n_{i\beta} n_{i\gamma''} ] [  \delta_{\beta' \gamma'''}  $\\$ + n_{i\beta'} n_{i\gamma'''}    ]T^{-1}_{\gamma''\gamma}T^{-1}_{\gamma'''\gamma'}$\end{tabular}\\  \vspace{1mm}
  $J_{\beta\beta';\gamma\gamma'}^{ab} $ &  $-\frac{a^2S^2}{V_{uc}} \epsilon_{\alpha\beta\gamma}\sum\limits_{\langle i,j\rangle} D_{ij,\alpha }n_{i\beta'} n_{j\gamma'}e_{ia}e_{ib}$ \\  \vspace{1mm}
  $K_{\beta\beta';\gamma\gamma'}^D $ & $\frac{2S^2}{V_{uc}} \epsilon_{\alpha\beta\gamma}\sum\limits_{\langle i,j\rangle} D_{ij,\alpha }n_{i\beta'} n_{j\gamma'}$ \\  \vspace{1mm}
  ${H}_{D, \gamma } $ &  $ \frac{2aS^2}{V_{uc}} \epsilon_{\alpha\beta\gamma}\sum\limits_{\langle i,j\rangle} (n_{i\alpha}-n_{j\alpha})D_{ij,\beta }$  \\  \vspace{1mm}
  $C_{\alpha\beta \gamma}$ & $\frac{2a^2S^2}{V_{uc}}\frac{\partial J }{\partial a }\sum\limits_{\langle i,j\rangle}e_{i\alpha}e_{i\beta}(1-n_{i\alpha'}n_{j\alpha'})T^{-1}_{\gamma' \gamma }(n_{i\gamma'}+n_{j\gamma'})$ \\  \vspace{1mm} 
  $d_{i\alpha}$ & $ a \frac{\partial D^z }{\partial a }\delta_{z\alpha}+a\frac{\partial D_\parallel}{\partial a }\epsilon_{\alpha\beta\gamma}e_{i,\beta}\delta_{z\gamma} $ \\ \vspace{1mm} 
  ${D}^{(1)\gamma \gamma'}_{\beta\beta'}$ & $\frac{2S^2}{V_{uc} }\sum\limits_{\langle i,j\rangle}\epsilon_{\beta' \eta\eta'}d_{i\beta}e_{i\gamma}e_{i\gamma'}n_{i,\eta}n_{j,\eta'}$ \\\vspace{1mm}
  ${D}^{(2)\alpha'\beta'}_{\alpha\beta\gamma}$ & \begin{tabular}[t]{@{}c@{}}$\frac{2aS^2}{V_{uc}}\sum\limits_{\langle i,j\rangle}d_{i,\alpha}e_{i,\alpha'}e_{i,\beta'}\epsilon_{\beta \beta'\gamma' }[(n_{i\beta'}-n_{j\beta'})\delta_{\gamma'\gamma''}$ \\ $-n_{i\beta'}n_{j\gamma'}(n_{i\gamma''}+n_{j\gamma''})]T^{-1}_{\gamma''\gamma}$\end{tabular}\\
  \hline
  \end{tabular}
\end{table} 

\subsection{Exchange interaction}
Assuming that the exchange parameter $J_{ij} $ only depends on the distance $\norm{\mathbf{r}_i-\mathbf{r}_j}$ between the lattice sites, we can expand $J_{ij}$ to first order in the strain $\varepsilon_{\alpha\beta}$ and spin $\omega_{\alpha\beta}$ tensors, 
\begin{align}\label{eq:app_exchange_parameter}
  J_{ij}
    & = J(a) + a \left(\frac{\partial J }{\partial a }\right) {e}_{i,\alpha} \left(\varepsilon_{\alpha\beta} + \omega_{\alpha\beta}\right) {e}_{i,\beta}, 
\end{align}
with strain tensor $\varepsilon_{\alpha\beta} = \frac{1}{2} \left( \frac{\partial {u}_\alpha(\mathbf{r}_i)}{\partial r_\beta} + \frac{\partial {u}_\beta(\mathbf{r}_i)}{\partial r_\alpha} \right)$ and spin tensor $\omega_{\alpha\beta} = \frac{1}{2} \left( \frac{\partial {u}_\alpha(\mathbf{r}_i)}{\partial r_\beta} - \frac{\partial {u}_\beta(\mathbf{r}_i)}{\partial r_\alpha} \right)$. The spin tensor $\omega$ is antisymmetric and, as a result, does not contribute to the exchange parameter. We can see this by writing the antisymmetric spin tensor as $\omega_{\alpha\beta} = \epsilon_{\alpha\beta\gamma} \Omega_{\gamma}$, which yields 
\begin{align}
    {e}_{i,\alpha} \omega_{\alpha\beta}{e}_{i,\beta}
        &= \mathbf{\Omega}\cdot (\hat{\mathbf{e}}_{i}\times \hat{\mathbf{e}}_{i}) = 0 .
\end{align}
Hence, $J_{ij}$ can be written as in Eq.~\eqref{eq:exchange_parameter}. 
The contribution to the  exchange from unit cell $l$ is (see Fig. \ref{fig:unit_cell})
\begin{align}
  H^{(l)}_{ex, \text{tot}}
    & =J_{13}\ \mathbf{S}_1^{\mathbf{r}_1(l)} \cdot \left(\mathbf{S}_3^{\mathbf{r}_1(l)+\hat{\mathbf{e}}_1}+\mathbf{S}_3^{\mathbf{r}_1(l)-\hat{\mathbf{e}}_1}\right)\notag\\
    & \quad + J_{21}\ \mathbf{S}_2^{\mathbf{r}_2(l)} \cdot \left(\mathbf{S}_1^{\mathbf{r}_2(l)+\hat{\mathbf{e}}_2}+\mathbf{S}_1^{\mathbf{r}_2(l)-\hat{\mathbf{e}}_2}\right)\notag\\
    & \quad + J_{32}\ \mathbf{S}_3^{\mathbf{r}_3(l)} \cdot \left(\mathbf{S}_2^{\mathbf{r}_3(l)+\hat{\mathbf{e}}_3}+\mathbf{S}_2^{\mathbf{r}_3(l)-\hat{\mathbf{e}}_3}\right),
\end{align}
where $i,j \in \{1,2,3\}$ label the three spins in the unit cell and $\mathbf{r}_i(l)$ denotes the position of spin $i$ in unit cell $l$. We divide the contribution into a term containing the usual exchange interaction $H^{(l)}_{ex}$ and a term containing the strain induced interaction $H^{(l)}_{ex,s}$. A gradient expansion for the spins yields
\begin{align}\label{eq:gradient_expansion}
  \mathbf{S}_j^{\mathbf{r}_i(l)\pm \hat{\mathbf{e}}_i} \simeq \mathbf{S}_i^{\mathbf{r}_i(l)} \pm a (\hat{\mathbf{e}}_i\cdot \nabla) \mathbf{S}_j^{\mathbf{r}_i(l)}+\frac{a^2}{2} (\hat{\mathbf{e}}_i\cdot \nabla)^2 \mathbf{S}_j^{\mathbf{r}_i(l)},
\end{align} 
which along with Eq.~\eqref{eq:exchange_parameter} leads to
\begin{align}
  H^{(l)}_{ex}
    & = 2J(a) \mathbf{S}_1^{\mathbf{r}_1(l)} \cdot \mathbf{S}_3^{\mathbf{r}_1(l)}\notag \\
    & \quad +J(a) a^2 \left(\mathbf{S}_1^{\mathbf{r}_1(l)} \cdot (\hat{\mathbf{e}}_1\cdot \nabla)^2 \mathbf{S}_3^{\mathbf{r}_1(l)}\right)\notag\\
    & \quad + 2J(a) \mathbf{S}_2^{\mathbf{r}_2(l)} \cdot \mathbf{S}_1^{\mathbf{r}_2(l)}\notag \\
    & \quad +J(a) a^2 \left(\mathbf{S}_2^{\mathbf{r}_2(l)} \cdot (\hat{\mathbf{e}}_2\cdot \nabla)^2 \mathbf{S}_1^{\mathbf{r}_1(l)}\right)\notag\\
    & \quad + 2J(a) \mathbf{S}_3^{\mathbf{r}_3(l)} \cdot \mathbf{S}_2^{\mathbf{r}_3(l)}\notag \\
    & \quad + J(a) a^2 \left(\mathbf{S}_3^{\mathbf{r}_3(l)} \cdot (\hat{\mathbf{e}}_3\cdot \nabla)^2 \mathbf{S}_2^{\mathbf{r}_3(l)}\right),\\
  H^{(l)}_{ex, s}  
    & = 2 A^{\alpha\beta}_{13}\ \mathbf{S}_1^{\mathbf{r}_1(l)} \cdot \mathbf{S}_3^{\mathbf{r}_1(l)}\ \varepsilon_{\alpha\beta}\notag\\
    & \quad + A_{13}^{\alpha\beta} a^2 \left(\mathbf{S}_1^{\mathbf{r}_1(l)} \cdot (\hat{\mathbf{e}}_1\cdot \nabla)^2 \mathbf{S}_3^{\mathbf{r}_1(l)}\right)\varepsilon_{\alpha\beta}\notag\\
    & \quad + 2 A^{\alpha\beta}_{21}\ \mathbf{S}_2^{\mathbf{r}_2(l)} \cdot \mathbf{S}_1^{\mathbf{r}_2(l)}\ \varepsilon_{\alpha\beta}\notag\\
    & \quad + A_{21}^{\alpha\beta} a^2 \left(\mathbf{S}_2^{\mathbf{r}_2(l)} \cdot (\hat{\mathbf{e}}_2\cdot \nabla)^2 \mathbf{S}_1^{\mathbf{r}_2(l)}\right)\varepsilon_{\alpha\beta}\notag\\
    & \quad + 2 A^{\alpha\beta}_{32}\ \mathbf{S}_3^{\mathbf{r}_3(l)} \cdot \mathbf{S}_2^{\mathbf{r}_3(l)}\ \varepsilon_{\alpha\beta}\notag\\
    & \quad + A_{32}^{\alpha\beta} a^2 \left(\mathbf{S}_3^{\mathbf{r}_3(l)} \cdot (\hat{\mathbf{e}}_3\cdot \nabla)^2 \mathbf{S}_2^{\mathbf{r}_3(l)}\right)\varepsilon_{\alpha\beta}.
\end{align}
Using Eq.~\eqref{eq:spin_approximation} we then find
\begin{align}
  H^{(l)}_{ex}
    & = 9 a^2S^2 J \mathbf{m}^2 - V_{uc}\Lambda_{mm'}^{\alpha\beta}\left[(\partial_\alpha R^T)(\partial_\beta R)\right]_{mm'},\label{Eq:Hlex} \\
  H^{(l)}_{ex, s}
    & = V_{uc} C_{\alpha\beta \gamma}\varepsilon_{\alpha\beta} m_\gamma , \label{Eq:Hlexs}
\end{align}
up to second order in $\mathbf{m}$, space-time derivatives, and strain. 
The tensors $\Lambda_{mm'}^{\alpha\beta}$ and $C_{\alpha\beta \gamma}$ are defined in Table \ref{tab:2}. 
From Eqs.~\eqref{Eq:Hlex}-\eqref{Eq:Hlexs}, we obtain $\mathcal{F}_e= H^{(l)}_{ex}/V_{uc}$ while $H^{(l)}_{ex,s}/V_{uc}$ produces the first term in Eq.~\eqref{Eq:Fs}.
\subsection{DMI}
A similar expansion for the DMI parameters yields
\begin{align}
  D^z_{ij}
    & = D^z(a) + a \left(\frac{\partial D^z }{\partial a }\right) {e}_{i,\alpha} \varepsilon_{\alpha\beta} {e}_{i,\beta},\\
  D^\parallel_{ij}
    & = D^\parallel(a) + a \left(\frac{\partial D^\parallel }{\partial a }\right) {e}_{i,\alpha} \varepsilon_{\alpha\beta} {e}_{i,\beta}.
\end{align}
The total DMI energy contribution from unit cell $l$ is
\begin{align}
  H^{(l)}_{DMI, \text{tot}}
    & =  \mathbf{D}_{13}\cdot\left[\mathbf{S}_1^{\mathbf{r}_1(l)} \times (\mathbf{S}_3^{\mathbf{r}_1(l)+\hat{\mathbf{e}}_1} + \mathbf{S}_3^{l-\hat{\mathbf{e}}_1}) \right]\notag\\
    & \quad + \mathbf{D}_{21}\cdot\left[\mathbf{S}_2^{\mathbf{r}_2(l)} \times (\mathbf{S}_1^{\mathbf{r}_2(l)+\hat{\mathbf{e}}_2} + \mathbf{S}_1^{\mathbf{r}_2(l)-\hat{\mathbf{e}}_2}) \right]\notag\\
    & \quad + \mathbf{D}_{32}\cdot\left[\mathbf{S}_3^{\mathbf{r}_3(l)} \times (\mathbf{S}_2^{\mathbf{r}_3(l)+\hat{\mathbf{e}}_3} + \mathbf{S}_2^{\mathbf{r}_3(l)-\hat{\mathbf{e}}_3}) \right].
\end{align}
Using the gradient expansion in Eq.~\eqref{eq:gradient_expansion} along with the expression for the spin in Eq.~\eqref{eq:spin_approximation},  we obtain
\begin{align}
  H^{(l)}_{DMI}
    & = V_{uc}(K_{\beta\beta';\gamma\gamma'}^DR_{\beta\beta'}R_{\gamma\gamma'}- \mathbf{H}_D\cdot \mathbf{m}\notag \\
    & \quad + J_{\beta\beta';\gamma\gamma'}^{ab}(\partial_a R_{\beta\beta'})(\partial_b R_{\gamma\gamma'}) );\\
  H^{(l)}_{DMI,s}
    & = V_{uc} ({D}^{(1)\alpha'\beta'}_{\alpha\beta}\varepsilon_{\alpha'\beta'}  R_{\alpha\beta} +{D}^{(2)\alpha'\beta'}_{\alpha\beta\gamma}\varepsilon_{\alpha'\beta'} R_{\alpha\beta} m_\gamma ),
\end{align}
to second order in $\mathbf{m}$, space-time derivatives, and strain. The tensors and $\mathbf{H}_D$ are defined in Table \ref{tab:2}. 
Consequently, $\mathcal{F}_{DMI}= H^{(l)}_{DMI}/V_{uc}$, whereas $H^{(l)}_{DMI,s}/V_{uc}$ yields the last two tems in Eq.~\eqref{Eq:Fs}. 
\subsection{Anisotropy}
The anisotropy energy per unit cell $l$ is
\begin{align}
  H^{(l)}_{an}
    & = \sum_{i=1,2,3} \left[K_\perp (\mathbf{S}_i\cdot \hat{\mathbf{z}})^2-K(\mathbf{S}_i\cdot \hat{\mathbf{n}}_i)^2\right].
\end{align}
The constants $K_\perp$ and $K$ describe the out-of-plane and in-plane anisotropies, respectively.  
In principle, strain reduces the system's symmetry, potentially producing new anisotropy terms.
As the primary purpose of this work is to investigate how strain-induced modifications of the exchange interactions can trigger chiral switching, we neglect these terms in our study.  
Hence, we have
\begin{align}
  H^{(l)}_{an}
    & = -V_{uc} R_{\alpha\beta}R_{\alpha'\beta'} (K^{(1)}_{\alpha\beta\alpha'\beta'}+m_\gamma K^{(2)}_{\alpha\beta\gamma\alpha'\beta'})\notag\\
    & \quad - V_{uc} R_{\alpha\beta}R_{\alpha'\beta'}m_\gamma m_{\gamma'} K^{(3)}_{\alpha\beta\gamma\alpha'\beta'\gamma'},
\end{align}
to second order in $\mathbf{m}$. 
\subsection{Temperature effects}
The effects of temperature are included via a stochastic white noise field $\mathbf{h}_i$: 
\begin{align}
  H^{(l)}_{h}
    & = - \sum_{i=1,2,3} \mathbf{h}_i \cdot \mathbf{S}_i.
\end{align}
Since the white noise field originates microscopically from modulations in the local magnetic anisotropy and exchange interactions induced by lattice vibrations -- which vary on the length scale of a single unit cell -- we assume $\mathbf{h}_i$ to be uniform within each unit cell. This gives
\begin{align}
    \sum_{i=1,2,3} \mathbf{h} \cdot \mathbf{S}_i
    & = - 3 a S\ \mathbf{h} \cdot  \mathbf{R} \mathbf{m}.
\end{align}

\subsection{Kinetic energy}
The kinetic energy per unit cell is \cite{Dombre:PRB1989},
\begin{align}
  T^{(l)}
    & = \sum_{i=1,2,3} \hbar \int dt \mathbf{A}[\mathbf{S}_i]\cdot \partial_t \mathbf{S}_i,\label{Eq:Tuc}
\end{align}
where $\mathbf{A}[\mathbf{S}_i]$ is the Berry phase vector potential. We expand the vector potential to first order in $\mathbf{\Delta}_i$, which gives $\mathbf{A}[\mathbf{S}_i] \simeq \mathbf{A}[S \mathbf{R}\hat{\mathbf{n}}_i]+ \frac{\partial \mathbf{A}(S \mathbf{R}\hat{\mathbf{n}}_i ) }{\partial {S}_\alpha}(S\mathbf{R} \mathbf{\Delta}_i)_\alpha$. Using $\nabla \times A(\mathbf{S}_i)=\mathbf{S}_i/S$ and $\epsilon_{\alpha \beta \gamma} R_{\alpha \alpha'} R_{\beta \beta'} R_{\gamma \gamma'} = \epsilon_{\alpha' \beta' \gamma'}$ we find, $\sum_i \hbar {A}_\alpha [S_i] \dot{S}_{i,\alpha}\approx \sum_i \hbar S \left[ A_{\alpha} [\mathbf{R}\hat{\mathbf{n}}_k]  (\dot{\mathbf{R}}\hat{\mathbf{n}}_i)_{\alpha} + a \epsilon_{\alpha \beta \gamma} L_{\alpha} n_{i,\beta} (\mathbf{R}^T \dot{\mathbf{R}}\hat{\mathbf{n}}_i)_{\gamma} \right]$.   Substituting this into Eq.~\eqref{Eq:Tuc}, we find~\cite{Dombre:PRB1989}
\begin{align}
  T^{(l)}
    & = 3 a \hbar S \mathbf{m}\cdot \mathbf{V}; \ {V}_\alpha=-\frac{1}{2}\epsilon_{\alpha\beta\gamma}(R^{-1}\partial_tR)_{\beta\gamma}. 
\end{align}

\section{Action in terms of nautical angles}\label{sec:fullaction}
We represent the rotation matrix in terms of the nautical angles, $\mathbf{R}(\theta,\phi,\psi)=\mathbf{R}_z(\theta)\mathbf{R}_y(\phi)\mathbf{R}_x(\psi)$. 
$\mathcal{F}_e= K_e \mathbf{m}^2$ whereas the DMI energy becomes ($\tilde{\mathcal{F}}\equiv V_{uc}\mathcal{F}$ and $\tilde{\mathcal{T}}\equiv V_{uc}\mathcal{T}$ )
\begin{align}
  \tilde{\mathcal{F}}_{DMI}
    & = -(K_d + E_s^{(2)})\cos\phi\cos\psi(a^2 m_
    x^2 + a^2 m_y^2-1).
\end{align}
For the anisotropy energy, we find $\tilde{\mathcal{F}}_{a}=  \tilde{\mathcal{F}}_{a}^{\parallel} + \tilde{\mathcal{\mathcal{F}}}_{a}^{z}$ where 
\begin{widetext}
  \begin{align}
    \tilde{\mathcal{F}}_{a}^{z}
      & = \frac{3}{2} K_{\bot} S^2 \bigg( \sin^2\phi + \cos^2\phi \sin^2\psi + a^2 \left(3 \sin^2\phi + \cos^2\phi \sin^2\psi\right) m_x^2 + a^2 \left(\sin^2\phi + 3 \cos^2\phi \sin^2\psi\right) m_y^2 \notag\\
      & \quad + 2 a^2 \cos^2\phi \cos^2\psi \ m_z^2 - 2 a \sin 2\phi m_x (\sin\psi + a \sin\psi\ m_y + a \cos\psi\ m_z)\notag\\ 
      & \quad + 2 a m_y \big(\sin^2\phi - \cos^2\phi \sin^2\psi + a \cos^2\phi \sin 2\psi\ m_z\big) \bigg);\\
    \tilde{\mathcal{F}}_{a}^{\parallel}
      & = \frac{1}{16} K S^2 \Big[  16 \left( \cos\theta \cos\psi + \sin\theta \sin\phi \sin\psi + 2a \cos\phi \sin\theta m_x + a (\cos\psi \sin\theta \sin\phi - \cos\theta \sin\psi) m_z \right)^2  \notag\\
      & \quad + \big[ (\cos\theta \cos\psi + \sin\theta \sin\phi \sin\psi) (1 + \sqrt{3} a m_x - 3 a m_y) + \cos\phi \sin\theta (-a m_x + \sqrt{3} (1 + a m_y)) \notag \\
      & \quad - 2a (\cos\psi \sin\theta \sin\phi - \cos\theta \sin\psi) m_z -
      \sqrt{3} \big[ (\cos\psi \sin\theta - \cos\theta \sin\phi \sin\psi) (1 + \sqrt{3} a m_x - 3 a m_y) \notag \\
      & \quad -  \cos\theta \cos\phi (-a m_x + \sqrt{3} (1 + a m_y)) + 2 a (\cos\theta \cos\psi \sin\phi + \sin\theta \sin\psi) m_z \big] \big]^2 \notag \\
      & \quad + \big[ (\cos\theta \cos\psi + \sin\theta \sin\phi \sin\psi) (-1 + \sqrt{3} a m_x + 3 a m_y) + \cos\phi \sin\theta (a m_x + \sqrt{3} (1 + a m_y)) \notag \\
      & \quad + 2a (\cos\psi \sin\theta \sin\phi - \cos\theta \sin\psi) m_z -
      \sqrt{3} \big[ -(\cos\psi \sin\theta - \cos\theta \sin\phi \sin\psi) (-1 + \sqrt{3} a m_x + 3 a m_y) \notag \\
      & \quad + \cos\theta \cos\phi (a m_x + \sqrt{3} (1 + a m_y)) + 2a (\cos\theta \cos\psi \sin\phi + \sin\theta \sin\psi) m_z \big] \big]^2 \Big].
  \end{align}
\end{widetext}
The free energy due to the stochastic field becomes,
\begin{align}
  \tilde{\mathcal{F}}_{h}
    & = 3 a S \Big( (h_x \cos\theta \cos\phi + h_y \cos\phi \sin\theta - h_z \sin\phi) m_x \notag\\
    & \quad + (h_z \cos\phi \sin\psi + h_x (-\cos\psi \sin\theta + \cos\theta \sin\phi \sin\psi) \notag\\
    & \quad + h_y (\cos\theta \cos\psi + \sin\theta \sin\phi \sin\psi)) m_y \notag\\
    & \quad + (h_z \cos\phi \cos\psi + h_y (\cos\psi \sin\theta \sin\phi - \cos\theta \sin\psi) \notag\\
    & \quad + h_x (\cos\theta \cos\psi \sin\phi + \sin\theta \sin\psi)) m_z \Big),
\end{align}
and, finally, the kinetic energy is,
\begin{align}
  \tilde{\mathcal{T}}
    & = 3 a \hbar S \Big( m_y (\dot{\theta} \cos\phi \sin\psi + \dot{\phi}\cos\psi ) \notag\\
    & \quad + m_x (\dot{\psi}-\dot{\theta}\sin\phi) + m_z (\dot{\theta}\cos\phi \cos\psi  - \dot{\phi}\sin\psi )
     \Big).
\end{align}


\end{document}